\documentclass[aps,prx,superscriptaddress,amsfonts,amsmath,amssymb,showpacs,floatfix,reprint,longbibliography]{revtex4-2}

\usepackage{url}
\usepackage{bm}
\usepackage{graphicx}
\usepackage{amsmath}
\usepackage{amstext}
\usepackage{amssymb}
\usepackage{amsfonts}
\usepackage{amsbsy}
\usepackage{verbatim}
\usepackage{color}
\usepackage[colorlinks=true, urlcolor=blue, linkcolor=blue, citecolor=blue, pdftex]{hyperref}
\usepackage{multirow}
\usepackage{floatrow}
\usepackage{float}
\usepackage{gensymb}
\usepackage{textcomp}
\usepackage{enumitem}
\usepackage[version=3]{mhchem}
\usepackage{afterpage}
\usepackage{pbox}
\usepackage{makecell}
\usepackage{dsfont}
\usepackage{siunitx}
\usepackage{fancyhdr}
\usepackage{stackengine,wasysym,scalerel}
\usepackage{latexsym}
\usepackage{cancel}
\usepackage{array, multirow, bigdelim, makecell, booktabs}

\newcommand{\dd}{\mathrm{d}}
\renewcommand{\vec}{\mathbf}
\newcommand{\vk}{\vec{k}}

\newcommand{\vq}{\vec{q}}
\newcommand{\BL}{\bar{\Lambda}}

\date{\today}

\begin{document}

\title{Dimerization tendencies of the pyrochlore Heisenberg antiferromagnet: A functional renormalization group perspective}

\author{Max Hering}
\affiliation{Helmholtz-Zentrum Berlin f\"{u}r Materialien und Energie, Hahn-Meitner Platz 1, 14109 Berlin, Germany}
\affiliation{Dahlem Center for Complex Quantum Systems and Fachbereich Physik, Freie Universit\"at Berlin, 14195 Berlin, Germany}
\author{Vincent Noculak}
\affiliation{Helmholtz-Zentrum Berlin f\"{u}r Materialien und Energie, Hahn-Meitner Platz 1, 14109 Berlin, Germany}
\affiliation{Dahlem Center for Complex Quantum Systems and Fachbereich Physik, Freie Universit\"at Berlin, 14195 Berlin, Germany}
\author{Francesco Ferrari}
\affiliation{Institute f\"{u}r Theoretische  Physik, Goethe Universit\"{a}t Frankfurt, Max-von-Laue-Stra\ss e 1, 60438 Frankfurt am Main, Germany}
\author{Yasir Iqbal}
\affiliation{Department of Physics and Quantum Centers in Diamond and Emerging Materials (QuCenDiEM) group, Indian Institute of Technology Madras, Chennai 600036, India}
\author{Johannes Reuther}
\affiliation{Helmholtz-Zentrum Berlin f\"{u}r Materialien und Energie, Hahn-Meitner Platz 1, 14109 Berlin, Germany} 
\affiliation{Dahlem Center for Complex Quantum Systems and Fachbereich Physik, Freie Universit\"at Berlin, 14195 Berlin, Germany}

\begin{abstract}
We investigate the ground state properties of the spin-$1/2$ pyrochlore Heisenberg antiferromagnet using pseudofermion functional renormalization group techniques. The first part of our analysis is based on an enhanced parton mean-field approach which takes into account fluctuation effects from renormalized vertex functions. Our implementation of this technique extends earlier approaches and resolves technical difficulties associated with a diagrammatic overcounting. Using various parton ans\"atze for quantum spin liquids, dimerized and nematic states our results indicate a tendency for lattice symmetry breaking in the ground state. While overall quantum spin liquids seem unfavorable in this system, the recently proposed monopole state still shows the strongest support among all spin liquid ans\"atze that we have tested, which is further confirmed by our complementary variational Monte Carlo calculations. In the second part of our investigation, we probe lattice symmetry breaking more directly by applying the pseudofermion functional renormalization group to perturbed systems. Our results from this technique confirm that the system's ground state either exhibits broken $C_3$ rotation symmetry, or a combination of inversion and $C_3$ symmetry breaking.
\end{abstract}

\maketitle

\section{Introduction}
The pyrochlore network is a paradigmatic lattice to study the effects of magnetic frustration and it arouses ongoing interest which has spanned several decades. As a consequence of the unique geometry featuring corner-sharing tetrahedra, even the classical nearest-neighbor Ising model on the pyrochlore lattice is highly non-trivial due to its extensively degenerate ground state manifold. This gives rise to a classical spin liquid that has been observed in a number of materials referred to as {\it spin ice} compounds, most notably the titanates Ho$_2$Ti$_2$O$_7$ and Dy$_2$Ti$_2$O$_7$~\cite{bramwell01}. The associated physical phenomena ranging from residual entropies and pinch point singularities to monopole excitations are characteristic to this fascinating and multifaceted research field.

What is even more remarkable, when including quantum fluctuations in the classical nearest-neighbor Ising model on the pyrochlore lattice, e.g., via transverse spin interactions $S_i^xS_j^x+S_i^yS_j^y$, it can be shown perturbatively that the system realizes the iconic $U(1)$ quantum spin liquid with emergent gauge photons and fractionalized magnetic charges. The possible identification of these phenomena in magnetic materials, so-called {\it quantum spin ice} compounds, has become a particularly challenging but also rewarding research direction.

Given the rich physical phenomenology of (quantum) spin ice systems it is natural to ask about the ground state properties of spin models where the transverse terms are beyond the perturbative regime, such as the spin-1/2 nearest-neighbor pyrochlore Heisenberg antiferromagnet. Due to the strong magnetic frustration and the nearby $U(1)$ quantum spin liquid, it seems plausible that spin liquid behavior survives in the Heisenberg limit and, indeed, this model is considered as a prime candidate for realizing a quantum spin liquid. On the other hand, early works also point out the possibility of a dimerized ground state~\cite{Harris-1991,Isoda-1998,Koga-2001,Tsunetsugu-2001a,Tsunetsugu-2001b,Berg-2003,Tchernyshyov-2006,Moessner-2006}. Due to the inherent difficulties in treating frustrated quantum spin models (particularly in three dimensions) and the notorious lack of controlled and unbiased numerical approaches, these questions remain highly non-trivial and an ultimate answer currently cannot be given by a single method alone. Further incentive for studying the pyrochlore Heisenberg antiferromagnet comes from experimental progress in realizing such systems. Most notably, the recently synthesized spin-1/2 oxynitride pyrochlore compound Lu$_2$Mo$_2$O$_5$N$_2$ does not show any experimental indications of magnetic long-range order~\cite{clark14,iqbal17}.

On the theoretical front, two large-scale numerical studies have recently investigated the nearest-neighbor pyrochlore Heisenberg antiferromagnet and have both found indications for a dimerized ground state that spontaneously breaks lattice symmetries~\cite{Neupert2021,hagymasi21}. Using complementary state-of-the-art methods, such as many-variable variational Monte Carlo, exact diagonalization, and density-matrix renormalization group these results provide a rather compelling argument against quantum spin liquid behavior.  

Inspired by these findings, the present work adds another and again complementary perspective to the ground state properties of the pyrochlore Heisenberg antiferromagnet using the pseudofermion functional renormalization group (PFFRG) method. Our analysis goes significantly beyond a previous PFFRG study~\cite{Iqbal2019} as it investigates additional scenarios for spontaneous breaking of lattice symmetries and includes a recently developed self-consistent PFFRG-enhanced parton mean-field treatment~\cite{hering19}. Since the latter approach has been rarely applied so far, in the beginning the focus lies on introducing, discussing, and extending it. This technique first assumes that the system's low energy behavior is described by a quadratic fermionic parton theory as is expected for a quantum spin liquid~\cite{Wen2002}. The parameters of this theory, such as spinon hopping and pairing amplitudes are self-consistently determined within a Fock-type approach. As a crucial difference compared to a pure mean-field treatment, however, the bare exchange couplings entering the self-consistent equations are replaced by the renormalized ones from PFFRG, hence, adding fluctuations well beyond mean-field. This approach allows us to compare the RG behaviors of different mean-field ans\"atze for quantum spin liquids and identify the preferred one. As an improvement of this technique with respect to an earlier implementation~\cite{hering19}, we resolve a methodological subtlety that may lead to an overestimation of mean-field amplitudes. First, restricting to ans\"atze of previously proposed quantum spin liquid candidates for the pyrochlore Heisenberg antiferromagnet we identify the monopole-antimonopole chiral spin liquid state from Ref.~\cite{Kim2008} as the preferred one, even under the effect of fluctuations beyond mean-field. This finding is confirmed by our additional large-scale ($N=6912$-site) variational Monte Carlo calculations yielding a projected energy per site $E/J=-0.459402(6)$ which is lower compared to the energy $E/J=-0.457354(5)$ of the monopole flux state of Ref.~\cite{Burnell2009}. Furthermore, the PFFRG approach is not restricted to spin liquid states only, but the real-space configuration of fermionic hopping/pairing terms may also mimic possible types of dimerization patterns. Including such symmetry breaking configurations in our analysis, we observe a clear dominance of dimer orders over spin liquids, associated with a breaking of inversion and $C_3$ rotation symmetry, in agreement with Refs.~\cite{Neupert2021,hagymasi21}. To further confirm this result we pursue a more direct approach where we impose the dimerization patterns as small perturbations in the spin Hamiltonian and apply PFFRG for this modified system. We find large responses to perturbations that break both inversion and $C_3$ rotation symmetry together, while patterns which only break inversion but not the $C_3$ rotation symmetry are not supported. These results are in agreement with our PFFRG-enhanced mean-field treatment and further corroborate the findings in Refs.~\cite{Neupert2021,hagymasi21} such that, in total, we conclude that a dimer valence bond state or a nematic state constitute the most likely ground state scenario of the pyrochlore Heisenberg antiferromagnet.

The rest of the paper is structured as follows: In Section~\ref{sec:PFFRG}, we introduce the PFFRG approach and its variants which will be applied to the pyrochlore Heisenberg antiferromagnet. The results of our analysis are presented in Section~\ref{sec:Results}, where Section~\ref{sec:Results_partons} first discusses our findings of the PFFRG-enhanced mean-field treatment and compares them with results from a complementary variational Monte Carlo study. This is followed by a direct investigation of symmetry breaking patterns via PFFRG in Section~\ref{sec:Results_symm_breaking}. The paper ends with a discussion of the presented methods in Section~\ref{sec:discussion} and a conclusion in Section~\ref{sec:conclusion}.

\section{Methods}\label{sec:PFFRG}
Most of the methods applied in this paper are based on the PFFRG technique. In this section, we give an introduction into its standard formulation (Sec.~\ref{sec:PFFRG_standard}) and then discuss the two types of extensions which are used for our numerical investigations in Sec.~\ref{sec:Results}. Particularly, we put an emphasis on introducing our PFFRG-enhanced parton mean-field treatment in Sec.~\ref{sec:SelfConsistency}, followed by a brief description of our approach to directly investigate lattice symmetry breaking dimer patterns, see Sec.~\ref{sec:symm_break}.

\begin{figure*}[t]
\centering
\includegraphics[width=14cm]{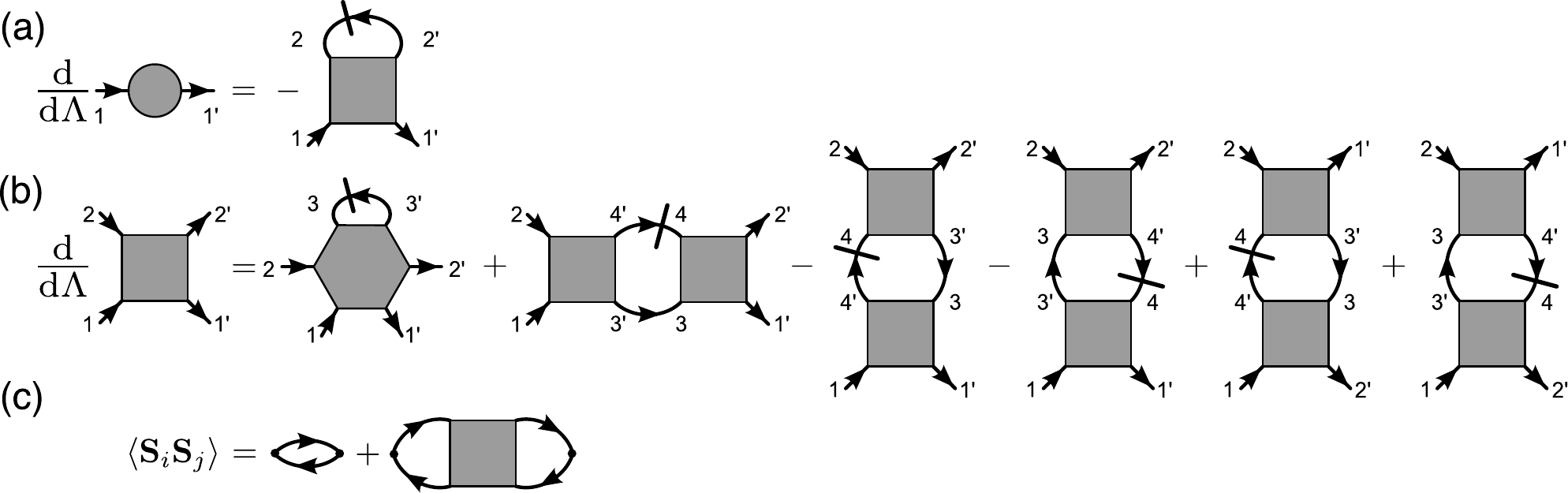}
\caption{(a), (b) PFFRG flow equations for $n$-particle vertex functions up to $n=2$: Regular (slashed) arrows represent the full Green's function $G^{\Lambda}$ (single-scale propagator $S^{\Lambda}$). The first equation in (a) couples the self energy $\Sigma^\Lambda$ (circle) to the $2$-particle vertex $\Gamma^\Lambda$ (square) and to itself via $S^{\Lambda}$. In the second equation (b), the $2$-particle vertex couples to the $3$-particle vertex (hexagon), itself, and the self energy. This procedure leads to an infinite hierarchy of coupled differential equations which we truncate by neglecting the $3$-particle vertex. In (c) the diagrammatic relation between the self energy, the $2$-particle vertex, and the spin-spin correlation $\langle\vec{S}_i\vec{S}_j\rangle$ is depicted.\label{fig:FRGeqs}}
\end{figure*}

\subsection{Standard PFFRG formulation}\label{sec:PFFRG_standard}
We first introduce the standard one-loop PFFRG method for quantum spin-$1/2$ systems~\cite{Reuther2010} which has proven to be a powerful tool for the investigation of magnetic properties of two and three dimensional spin systems with different types of interactions~\cite{Reuther2010,Reuther2011_2,Reuther-2011anisotropic,Reuther-2011honeycomb,Singh2012,Suttner2014,Iqbal2015b,Rousochatzakis2015_2,Iqbal2016,Iqbal-2016tri,Iqbal2016a,Buessen2016,Balz2016,Hering2017,Baez2017,iqbal17,keles18,keles18_2,Iqbal-2018spiral,Buessen2018a,Buessen2018b,Roscher2018,Rueck2018,Chillal2020,Ivica-2021,Iqbal2019,Buessen2019,Ghosh-2019bcc,Ghosh-2019pyro,Kiese2020,thoenniss20,Kiese2020_2,Iida-2020,Buessen2021,niggemann21}. Here, we apply it to a Heisenberg model on a pyrochlore lattice,
\begin{equation}
\label{eq:Hamiltonian}
    \mathcal{H}=\sum_{(i,j)}J_{ij} \, \vec{S}_i\cdot\vec{S}_j\;,
\end{equation}
where $(i,j)$ are pairs of sites and $J_{ij}$ is finite only on nearest-neighbor bonds; in this case $J_{ij}\equiv J>0$. The PFFRG treats this model by utilizing a pseudofermionic description of $S=1/2$ spin operators~\cite{Abrikosov1965,Abrikosov1970} also called parton representation,
\begin{equation}\label{eq:Arbikosov}
    S^{\mu}_{i}=\sum\limits_{\alpha, \beta}\frac{1}{2}f^{\dagger}_{i, \alpha}\sigma^{\mu}_{\alpha \beta}f_{i, \beta}, 
\end{equation}
where $f^{\dagger}_{i, \alpha}$ ($f_{i, \alpha}$) creates (annihilates) a fermion with spin $\alpha \in \{\uparrow, \, \downarrow\}$ at lattice site $i$ and $\sigma^{\mu}$ with $\mu \in \{x,\,y,\,z\}$ are the standard $2\times 2$ Pauli matrices. This rewriting enables one to employ common quantum-many-body techniques based on Feynman diagrams, including functional renormalization group methods~\cite{Wetterich1993}. Note, however, that the pseudofermion representation in Eq.~\eqref{eq:Arbikosov} artificially enlarges the Hilbert space by introducing two $S=0$ states per lattice site which may lead to artifacts in the numerical outcomes when pursuing the PFFRG procedure as explained below. Various previous works have discussed this effect~\cite{Baez2017,niggemann21,thoenniss20} and find that such artifacts are minor or not observable, at least as long as ground state properties are considered (as is also done in this work).

Since a quadratic spin Hamiltonian, rewritten in terms of pseudofermions is purely quartic in these auxiliary particles, the bare fermionic propagator in Matsubara space has the simple form $G_0(\omega)=\frac{1}{i \omega}$ and is strictly local in real space. The key manipulation within PFFRG is to regularize $G_0(\omega)$ via an infrared cutoff $\Lambda$:
\begin{equation}
    G_0(\omega)=\frac{1}{i \omega} \longrightarrow G^{\Lambda}_0(\omega)=\frac{\theta(|\omega|-\Lambda)}{i \omega}\;.
\end{equation}
This insertion leads to a cutoff-dependent generating functional for the one-particle-irreducible vertex functions. By taking the full cutoff derivative of these vertex functions, one arrives at an infinite hierarchy of coupled differential equations, the so-called \textit{flow equations}. Since the full set of equations is beyond numerical solvability, the hierarchy of equations needs to be truncated, which in our one-loop approach is done by neglecting the $3$-particle vertex. One then arrives at the flow equations for the fermionic self-energy written as $\Sigma^\Lambda(\omega)=-i\gamma^\Lambda(\omega)$ [where $\gamma(\omega)$ is real valued] and the $2$-particle vertex $\Gamma^\Lambda_{ij}(s,t,u)$ depicted diagrammatically in Figs.~\ref{fig:FRGeqs}(a) and~\ref{fig:FRGeqs}(b). Note that the three Matsubara frequency arguments of $\Gamma^\Lambda$ take into account energy conservation and are defined in a way that the vertex on the left hand side of Fig.~\ref{fig:FRGeqs}(b) corresponds to $\Gamma_{ij}^\Lambda(\omega_{1'}+\omega_{2'},\omega_{1'}-\omega_1,\omega_{1'}-\omega_2)$. Here, the frequencies $\omega_1$, $\omega_1'$, $\ldots$ are the ones on the external fermion lines labelled accordingly. A crucial technical necessity is the implementation of the so-called Katanin scheme~\cite{Katanin2004,Salmhofer2004} where the single-scale propagator [slashed line in Fig.~\ref{fig:FRGeqs}(b)] in the flow equation for $\Gamma^\Lambda$ is given by the full derivative of the renormalized propagator, $S^{\Lambda}(\omega)=-\frac{\dd}{\dd\Lambda}G^{\Lambda}(\omega)$. This insertion re-includes certain contributions of the $3$-particle vertex and ensures that quantum fluctuations are included on a level that captures the subtle interplay of ordering and disordering tendencies in quantum spin systems.

Apart from the aforementioned truncation, two more approximations are required to yield a finite set of differential equations in the zero-temperature limit. First, the Matsubara-frequencies, which at $T=0$ are continuous variables, need to be replaced by a discrete mesh, in our case consisting of 64 frequencies. Second, all $2$-particle vertices which exceed a given real-space distance must be neglected. In Sec.~\ref{sec:Results_partons} (Sec.~\ref{sec:Results_symm_breaking}), we truncate correlations after $10$ nearest-neighbor bonds (outside a sphere with a radius of $5$ nearest-neighbor distances). This corresponds to a finite size approximation that accounts for correlations in a cluster of $741$ ($381$) sites.
 
In its standard formulation, the resulting flow equations are numerically solved, starting in the limit $\Lambda\rightarrow\infty$ where the $2$-particle vertex is given by the bare interactions, $\Gamma^{\Lambda\rightarrow\infty}_{ij}\equiv J_{ij}$. While lowering $\Lambda$, magnetic instabilities may be detected by a breakdown of the RG flow of the $2$-particle vertex, or otherwise, the flow continues down to $\Lambda=0$ indicating non-magnetic ground state behavior. Having reached this physical cutoff-free limit, the real-space spin-spin correlations $\langle\vec{S}_i\vec{S}_j\rangle$ (or their Fourier-transform into momentum space), which depend on the $2$-particle vertex as shown in Fig.~\ref{fig:FRGeqs}(c) may be further investigated. This type of analysis has already been performed for the nearest-neighbor pyrochlore Heisenberg antiferromagnet in Ref.~\cite{Iqbal2019} demonstrating the absence of magnetic long-range order. The approach outlined in the next section, which will be applied in Sec.~\ref{sec:Results}, pursues a different strategy. There, the $2$-particle vertex is not used as a direct diagnostic tool but rather forms the basis for a more involved post-processing analysis.

\subsection{PFFRG-enhanced parton mean-field approach}\label{sec:SelfConsistency}

\begin{figure}[t]
\centering
\includegraphics[width=8.5cm]{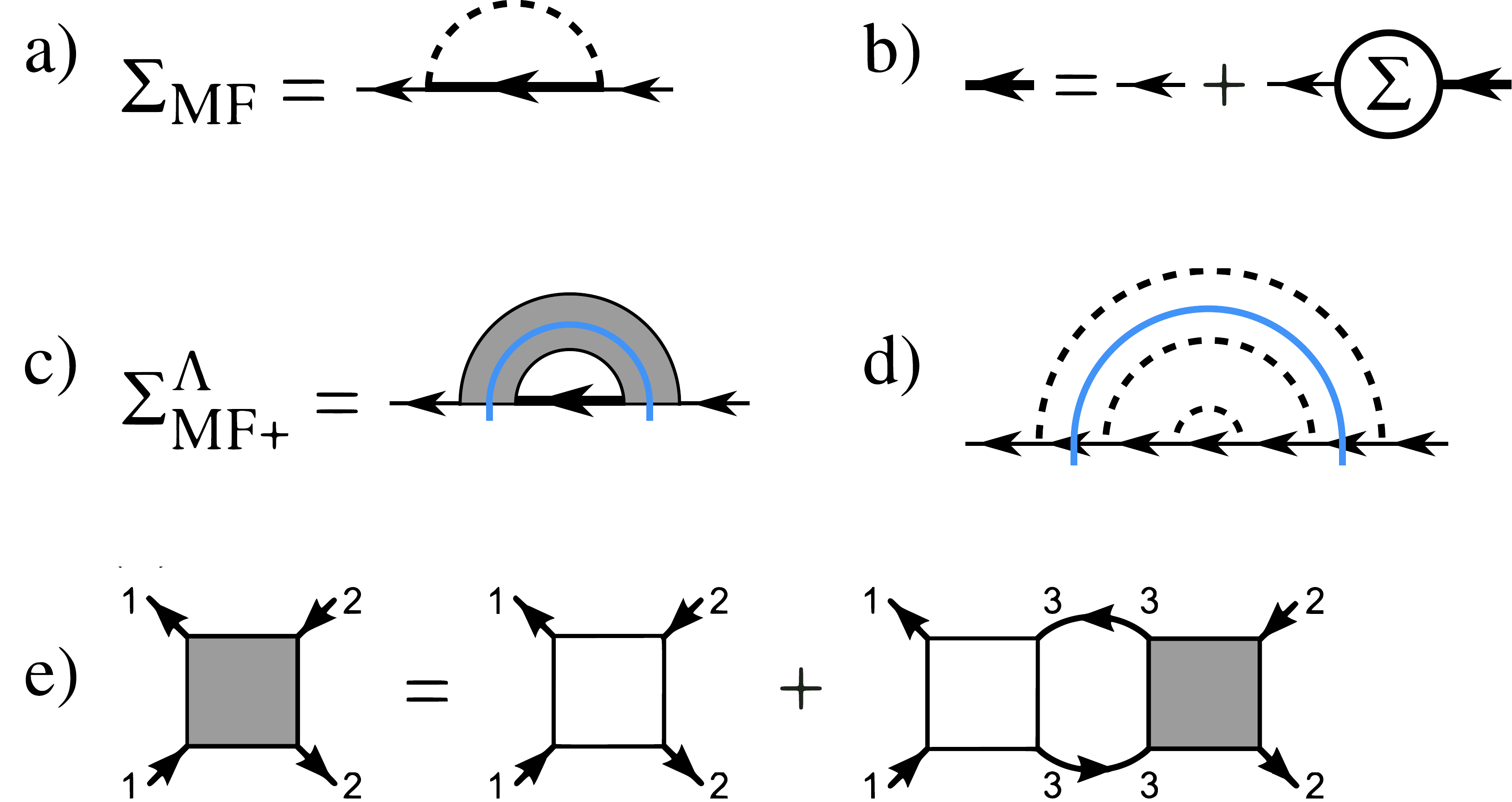}
\caption{Diagrammatic representations of different self-consistent schemes. (a) Bare self-consistent Fock mean-field approach. The dashed line is a bare interaction $\sim J_{ij}$. The thick line is a dressed propagator resulting from Dyson's equation in (b). The thin line in (b) is the free fermionic propagator $G_{0}(\omega)=\frac{1}{i\omega}$. In an enhanced parton mean-field scheme, Dyson's equation is evaluated together with the self-energy $\Sigma_{\text{MF+}}$ shown in (c) making use of the renormalized vertex from PFFRG (gray box). Note that in this approximation, the thin line in (b) also contains the self-energy $\gamma(\omega)$ from PFFRG. (d) Illustration of a contribution from the bare self-consistent Fock mean-field scheme, obtained via iteration, see main text for details. Blue lines in (c) and (d) illustrate a cut for a two-particle decoupling in the crossed particle-hole channel. To avoid overcounting in $\Sigma_{\text{MF+}}$, the enhanced mean-field equation in (c) needs to be evaluated with a two-particle irreducible vertex $\tilde{\Gamma}^\Lambda$ (white square) which follows from the Bethe-Salpeter equation in (e). Note that in this equation the propagator line is dressed with $\gamma^\Lambda$ but not with $\Sigma_{\text{MF+}}$. \label{fig:BSE}}
\end{figure}

The approach presented in this section is directly adapted to the states it aims to describe, namely non-magnetic spin states and primarily quantum spin liquids. Therefore, we first give a brief introduction into the general phenomenology and low-energy behavior of quantum spin liquids.

A characteristic property of a quantum spin liquid is that it features fractional and spinful quasiparticles called spinons which we here assume to be fermions. The fractional property stems from the fact that it requires two spinons to create a conventional $\Delta S=\pm1$ excitation. Since this property is inherently given by the fermionic parton representation of spin operators in Eq.~\eqref{eq:Arbikosov} it is natural to construct a low-energy effective theory for quantum spin liquids based on this rewriting. Furthermore, in quantum spin liquids the spinons described by the fermionic partons $f_{i,\alpha}$ are free in the sense that they do not experience long-range confining forces between them. Hence, at low energies the situation can be described by a general quadratic model in $f_{i,\alpha}$, including spinon hopping $\chi_{ij}$ and pairing $\eta_{ij}$,
\begin{equation}\label{eq:ham_mf}
\mathcal{H}_{\text{MF}}=-\frac{3}{8}\sum_{i,j} \left(\chi_{ij}f_{j,\alpha}^\dagger f_{i,\beta} \delta_{\alpha\beta}+\eta_{ij}f_{i,\alpha}^\dagger f_{j,\beta}^\dagger\epsilon_{\alpha\beta}+\text{H.c.}\right)\,.
\end{equation}
Here, the sums over spin indices are implicit and the anti-symmetric tensor $\epsilon$ has vanishing diagonal entries and the off-diagonals $\epsilon_{\uparrow \downarrow}=-\epsilon_{\downarrow \uparrow }=1$. Furthermore, the identities $\chi^{\dagger}_{ij}=\chi_{ji}$ and $\eta_{ij}=\eta_{ji}$ must hold. Due to the spin-isotropy of Heisenberg interactions, Eq.~\eqref{eq:ham_mf} only contains spin-rotation symmetric terms. 

It is clear that Eq.~\eqref{eq:ham_mf} alone cannot describe quantum spin liquids since its eigenstates generally contain contributions from the unphysical Hilbert space sectors and are thus not even proper spin states. To resolve this problem one introduces gauge fluctuations in the amplitudes $\chi_{ij}$ and $\eta_{ij}$ which in the simplest case of a $\mathds{Z}_2$ gauge theory correspond to sign fluctuations $\chi_{ij}\rightarrow\pm\chi_{ij}$, $\eta_{ij}\rightarrow\pm\eta_{ij}$. The crucial property of a $\mathds{Z}_2$ gauge theory is that excitations in the gauge field (so called visons) are gapped such that in the low-energy limit even the bare quadratic theory in Eq.~\eqref{eq:ham_mf} without gauge fluctuations provides a faithful description of quantum spin liquids.

The fact that the original strongly interacting spin model that is quartic in the fermions $f_{i,\alpha}$ becomes effectively quadratic at low energies implies that a mean-field decoupling in the hopping and pairing channels describes the system reasonably well [hence, the index `MF' in Eq.~\eqref{eq:ham_mf}]. This means that on this simple level of approximation the amplitudes $\chi_{ij}$ and $\eta_{ij}$ can be thought of as resulting from the self-consistency conditions
\begin{equation}\label{eq:chi_eta}
    \chi_{ij}= J_{ij}\langle f^{\dagger}_{i,\alpha} f_{j,\alpha}  \rangle\;,\quad\eta_{ij}=J_{ij}\langle  f_{j,\alpha} f_{i,\beta}\epsilon_{\beta \alpha} \rangle\;.
\end{equation}
Translating these conditions into Feynman diagrams, $\chi_{ij}$ and $\eta_{ij}$ have the form of a self-energy $\Sigma_{\text{MF}}$ which consists of a convolution of an equal-time fermionic propagator and a bare interaction line, as shown in Fig.~\ref{fig:BSE}(a). More precisely, to incorporate fermionic pairing, self-energies and propagators need to be extended to a $2\times2$ structure in Nambu space where
\begin{equation}\label{eq:UMatrix}
\Sigma_{\text{MF}} =  u_{ij} = \left (\begin{array}{cc}
       \chi^{\dagger}_{ij} & \eta_{ij}  \\
       \eta^{\dagger}_{ij} & -\chi_{ij}
   \end{array} \right) \, .
\end{equation}
To close self-consistency within the diagrammatic framework, the self-energy is fed back into Dyson's equation shown in Fig.~\ref{fig:BSE}(b), which in total corresponds to a Fock mean-field decoupling. In terms of explicit expressions in momentum space the self-consistent equation for $u_{\vk}$ reads
\begin{equation}\label{eq:FRGDecoupling0}
    u_{\vk}=-\int\limits^{\infty}_{-\infty} \frac{\dd \omega}{2\pi} \int\limits_{\mathrm{BZ}} \frac{\dd\vec{q}}{V_\text{BZ}}  J_{\vk-\vec{q}}\left[G_0^{-1}(\omega)-u_\vec{q}\right]^{-1}\,.
\end{equation}
For non-Bravais lattices like the pyrochlore lattice, this equation must be interpreted as a matrix equation in sublattice space. The momentum integral is carried out over a single Brillouin zone with volume $V_\text{BZ}$.

On this level, Eq.~\eqref{eq:FRGDecoupling0} and Figs.~\ref{fig:BSE}(a) and ~\ref{fig:BSE}(b) correspond to a standard parton mean-field theory for quantum spin liquids, expressed in terms of Feynman diagrams. In this technique different ans\"atze for $u_{ij}$ may be tested with respect to non-vanishing solutions and the one with the smallest mean-field free energy gives an indication of the system's spin liquid ground state.

We now explain our extension of this approach which we call `PFFRG-enhanced parton mean-field approach'. The key step is to replace the bare Heisenberg couplings $J_{ij}$ and the free fermionic propagator $G_0(\omega)=\frac{1}{i\omega}$ in Eq.~\eqref{eq:FRGDecoupling0} by the corresponding renormalized quantities from PFFRG,
\begin{equation}
    J_{ij}\rightarrow\Gamma_{ij}^\Lambda(s,t,u)\;,\quad G_0(\omega)\rightarrow \frac{1}{i\omega+i\gamma^\Lambda(\omega)}\;,
\end{equation}
including their full real-space and frequency dependence. The corresponding improved self-energy, called $\Sigma_{\text{MF+}}^\Lambda$, is depicted diagrammatically in Fig.~\ref{fig:BSE}(c) and the explicit self-consistent equation reads
\begin{align}\label{eq:self_conistent_old}
     u^{\bar{\Lambda}}_{\vk}&=-\int\limits^{\infty}_{-\infty} \frac{\dd \omega}{2\pi} \int\limits_{\mathrm{BZ}} \frac{\dd\vec{q}}{V_\text{BZ}} \Gamma^{\bar{\Lambda}}_{\vk-\vec{q}}(\omega,-\omega,0) \nonumber \\ & \times\left[i\omega+i\gamma^{\bar{\Lambda}}(\omega) - u^{\bar{\Lambda}}_{\vec{q}} \right]^{-1}\;.   
\end{align}
Since the renormalized vertices $\gamma^\Lambda$ and $\Gamma^\Lambda$ depend on $\Lambda$ there is the freedom to choose $\Lambda$ for which Eq.~\eqref{eq:self_conistent_old} is evaluated. We call this cutoff parameter the `decoupling scale' $\bar{\Lambda}$ to distinguish it from the renormalization group scale $\Lambda$ used in the PFFRG flow equations of Fig.~\ref{fig:FRGeqs}. By varying $\bar{\Lambda}$ from $\bar{\Lambda}=\infty$ to $\bar{\Lambda}=0$ one can smoothly interpolate between a bare mean-field scheme and an approach that uses the fully one-loop renormalized vertices. Hence, Eq.~\eqref{eq:self_conistent_old} extends the method well beyond mean-field since the vertices from PFFRG include additional dynamics and real-space dependencies ($\Gamma_{ij}^\Lambda$ become spatially more spread-out than $J_{ij}$) not contained in $J_{ij}$. Effectively, this can be thought of as taking into account gauge fluctuations and interactions between partons which are not contained in Eq.~\eqref{eq:ham_mf}. We therefore expect that the obtained spinon parameters $\chi_{ij}$ and $\eta_{ij}$ provide a better approximation of the system's low energy behavior.

This general type of combined FRG plus mean-field treatment has first been applied in Ref.~\cite{Reiss2007} to investigate competing instabilities in a two-dimensional Hubbard model. Since the mean-field Hamiltonian may explicitly break the symmetries of our system, an important benefit is that ordered phases can be directly accessed which is not easily possible within FRG alone. The extension to quantum spin systems in the pseudofermion representation has later been formulated in Ref.~\cite{hering19}. This latter work discusses solutions of Eq.~\eqref{eq:self_conistent_old} to characterize spin liquid phases in square and kagome Heisenberg antiferromagnets. It is worth emphasizing that for the investigation of quantum spin liquids, the objective of the approach differs from Ref.~\cite{Reiss2007} in the sense that the possibility of accessing symmetry broken phases is not exploited. Rather, Eq.~\eqref{eq:self_conistent_old} is solved for different ans\"atze $u_{ij}$ to identify the one that is realized at low energies.

Here, we develop the method of Eq.~\eqref{eq:self_conistent_old} and Ref.~\cite{hering19} further, by resolving a technical difficulty associated with an overcounting of diagrams that has first been noticed in Ref.~\cite{Wang2014} (there again in the context of a two-dimensional Hubbard model). To illustrate the overcounting, let us examine an iterative solution of the bare mean-field scheme in Fig.~\ref{fig:BSE}(a). A particular class of diagrams contributing to the solution has the form of the nested graphs in Fig.~\ref{fig:BSE}(d). These diagrams have the property that they decompose into disconnected graphs when cutting the two propagators along the blue line. Now we upgrade the bare interaction lines $\sim J$ to renormalized two-particle vertices $\Gamma^\Lambda$ from PFFRG. It is clear that if the vertex $\Gamma^\Lambda$ has the same property of decomposing into disconnected diagrams upon cutting two propagators in the crossed particle-hole channel, this creates an overcounting of terms (which already occurs in second order in $J$). The contributions of $\Gamma^\Lambda$ with this property have the structure of particle-hole ladder diagrams. To avoid this problem only those contributions $\tilde{\Gamma}^\Lambda$ of $\Gamma^\Lambda$ should be considered which are two-particle irreducible in the crossed particle-hole channel. Isolating the contributions $\tilde{\Gamma}^\Lambda$ corresponds to solving the Bethe-Salpeter equation (see Appendix~\ref{app:BSE}) in the crossed particle-hole channel as depicted in Fig.~\ref{fig:BSE}(e) and explicitly given by
\begin{align}\label{eq:BetheSalpeter}
    \Gamma^{\Lambda}_{\vk-\vq}&(\omega,-\omega,0)=\tilde{\Gamma}^{\Lambda}_{\vk-\vq}(\omega,-\omega,0) \nonumber \\ -&\int\limits^{\infty}_{-\infty} \frac{\dd \omega'}{2\pi} \int\limits_{\mathrm{BZ}} \frac{\dd\vec{p}}{V_\text{BZ}}\tilde{\Gamma}^{\Lambda}_{\vk-\vec{p}}(\omega',-\omega',0) \nonumber \\ &\times\Gamma^{\Lambda}_{\vec{p}-\vec{q}}(\omega'+\omega,\omega'-\omega,0)\left(G^{\Lambda}(\omega')\right)^2 \, .
\end{align}
Additionally, it needs to be ensured that $\Sigma_{\text{MF+}}^\Lambda$ only contains diagrammatic contributions that actually depend on the spinon amplitudes $\chi_{ij}$ and $\eta_{ij}$. Other terms independent of these parameters are local in real space and imaginary and, therefore, contribute to $\gamma(\omega)$. These latter self-energy terms are, however, already generated within PFFRG such that including them would lead to another source of overcounting. To overcome this problem, we additionally subtract the zeroth order term in $u_{ij}$ on the right hand side of Eq.~\eqref{eq:self_conistent_old}. Based on these considerations we can now formulate the corrected and final self-consistent equation containing $\tilde{\Gamma}^\Lambda$ instead of $\Gamma^\Lambda$:
\begin{align}\label{eq:FRGDecoupling}
    u^{\bar{\Lambda}}_{\vk}&=-\int\limits^{\infty}_{-\infty} \frac{\dd \omega}{2\pi} \int\limits_{\mathrm{BZ}} \frac{\dd\vec{q}}{V_{\mathrm{BZ}}} \tilde{\Gamma}^{\bar{\Lambda}}_{\vk-\vec{q}}(\omega,-\omega,0) \nonumber \\ &\times\left \{ \left[i\omega+i\gamma^{\bar{\Lambda}}(\omega) - u^{\bar{\Lambda}}_{\vec{q}} \right]^{-1} -\left[i\omega+i\gamma^{\bar{\Lambda}}(\omega)\right]^{-1} \right \}\;.
\end{align}
See Appendix~\ref{app:SCSol} for numerical details about how we solve this equation. The error in Eq.~\eqref{eq:self_conistent_old} resulting from overcounting may in general be drastic: If the vertex $\Gamma^\Lambda$ diverges during the RG flow due to an instability this will feed back into the self-consistent equation, leading to unphysical divergent amplitudes $u_{ij}$. In our specific situation where the $2$-particle vertex does not display any instabilities during the RG flow, the consequences of overcounting are less disastrous and solely have a quantitative effect. We base the following analysis on the more accurate self-consistent scheme in Eq.~\eqref{eq:FRGDecoupling}.

An unconstrained investigation of Eq.~\eqref{eq:FRGDecoupling} where $\chi_{ij}$ and $\eta_{ij}$ are taken as free parameters on all bonds $(i,j)$, is generally too complicated to be performed numerically. Hence, an ansatz for $u_{ij}$ is made which consists of a small number of free variables (typically amplitudes $|\chi_{ij}|$ and $|\eta_{ij}|$ while their phase relations on symmetry-related bonds are fixed) which are then self-consistently determined. Despite this reduction of complexity, there still exist large ansatz classes. For example, $u_{ij}$ can also become finite on bonds $(i,j)$ where $J_{ij}=0$, if $\Gamma^\Lambda_{ij}\neq0$. However, such a spread of amplitudes $u_{ij}$ in real space beyond the range of exchange couplings is typically a small effect, such that in our investigation below we neglect all spinon amplitudes beyond nearest neighbors. Moreover, the ans\"atze $u_{ij}$ do not need to obey the symmetries of the underlying lattice, even for symmetric quantum spin liquids. More precisely, possible ans\"atze are taken from a projective symmetry group analysis which imposes a weaker condition on the symmetries of $u_{ij}$, according to which a gauge transformation must exist such that the combined application of the symmetry transformation and the gauge operation leaves the ansatz invariant (so-called {\it projective} implementation of symmetries)~\cite{Wen2002,liu21}. One may also investigate states where lattice symmetries are explicitly (\textit{i.e.}, even projectively) broken such as for dimer valence bond solids. In the corresponding ans\"atze, $\chi_{ij}$ is taken to be finite only on bonds which are occupied by a dimer and zero otherwise (furthermore, $\eta_{ij}=0$ on all bonds)
~\cite{Affleck1988,Marston1989}. This type of ansatz reflects the physical properties of a dimer state where on length scales beyond the extent of singlet dimers spinons are no longer free quasiparticles, \textit{e.g.}, they can only hop within a dimer but not between dimers.

Having calculated the amplitudes $u_{ij}$ self-consistently for different ans\"atze, one needs a criterion that indicates which one describes the system's low energy physics best. In an ideal situation where the free energy functional for these amplitudes would be known, one could identify the ansatz which minimizes the free energy. However, within FRG the free energy functional including effects of renormalization beyond mean-field is not easily accessible (even though first approaches have accomplished parts of this task~\cite{niggemann21}). Therefore, we use a more basic approach and simply compare the sizes of the self-consistently calculated amplitudes $|\chi_{ij}|$ and $|\eta_{ij}|$ for different ans\"atze aiming to find the largest ones (note that for simplicity each ansatz investigated below is only characterized by one free parameter). This procedure is well justified from a pure mean-field perspective, where the energy expectation value $E_{\text{MF}}$ of Eq.~\eqref{eq:ham_mf} is given by
\begin{equation}\label{eq:EMF}
E_{\text{MF}}=\langle\mathcal{H}_\text{MF}\rangle\sim-\sum_{ij}J_{ij}^{-1}\left(|\chi_{ij}|^2+|\eta_{ij}|^2\right)\;,    
\end{equation}
which is minimized by the ansatz with the largest amplitudes.

\subsection{Investigation of symmetry breaking perturbations}\label{sec:symm_break}
To complement the analysis of the PFFRG-enhanced parton approach we use another and more straightforward method to directly probe the system's ground state with respect to symmetry breaking orders such as dimerization. This amounts to applying the standard PFFRG technique from Sec.~\ref{sec:PFFRG_standard}, but with slightly modified coupling constants $J$ mimicking the dimer pattern. Particularly, we start with a set of bonds $(i,j)$ which initially all carry the same interactions $J$ (such as nearest-neighbor bonds). We then partition these bonds into two groups $B^+$ and $B^-$ in a way that the bonds $B^+$ are those carrying a dimer and perturb the couplings $J$ by a small parameter $\delta>0$ as follows:
\begin{align}
    &(i,j)\in B^+:\;J_{ij}\longrightarrow J_{ij}+\delta\;,\notag\\
    &(i,j)\in B^-:\;J_{ij}\longrightarrow J_{ij}-\delta\;.
\end{align}
Running PFFRG for this system we keep track of the static spin-spin correlations
\begin{equation}
    \chi_{ij}^\Lambda=\int_0^\infty \dd \tau \, \langle\vec{S}_i(0)\vec{S}_j(\tau)\rangle
\end{equation}
[see Fig.~\ref{fig:FRGeqs}(c)] for all weakened and strengthened bonds. Of particular interest is the dimer response function $\chi_{D,ijkl}^\Lambda$ which measures how strongly this perturbation affects the correlations:
\begin{equation}\label{eq:dimer_response}
    \chi_{D,ijkl}^\Lambda=\frac{J}{\delta}\frac{|\chi_{ij}^\Lambda-\chi_{kl}^\Lambda|}{\chi_{ij}^\Lambda+\chi_{kl}^\Lambda}\;,\text{where}\;(i,j)\in B^+,\;(k,l)\in B^-\;.
\end{equation}
This quantity is defined such that at the beginning of the RG flow $\chi_{D,ijkl}^{\Lambda=\infty}=1$. If the dimer response function grows towards values much larger than one as $\Lambda$ approaches the cutoff-free limit ($\chi_{D,ijkl}^{\Lambda\rightarrow0}\gg1$) this indicates the tendency for dimerization. Furthermore, the size of the response function can be compared for different patterns to identify the preferred one. Note that for a given symmetry breaking pattern, \textit{i.e.}, for a certain partitioning of bonds into $B^+$ and $B^-$, the quantity $\chi_{D,ijkl}^\Lambda$ may not be uniquely given, but depends on the precise choice of strengthened/weakened bonds $(i,j)$ and $(k,l)$ for which the spin-correlations are compared to each other. This method cannot only be applied to dimer patterns but to all other types of lattice symmetry breaking states such as nematic order while, on the other hand, probing quantum spin liquid behavior directly is not possible (at least not beyond the observation that lattice symmetry breaking is absent).

\section{Results}\label{sec:Results}
\subsection{PFFRG-enhanced parton mean-field theory}\label{sec:Results_partons}

\begin{figure*}[t]
\centering
\includegraphics[width=17cm]{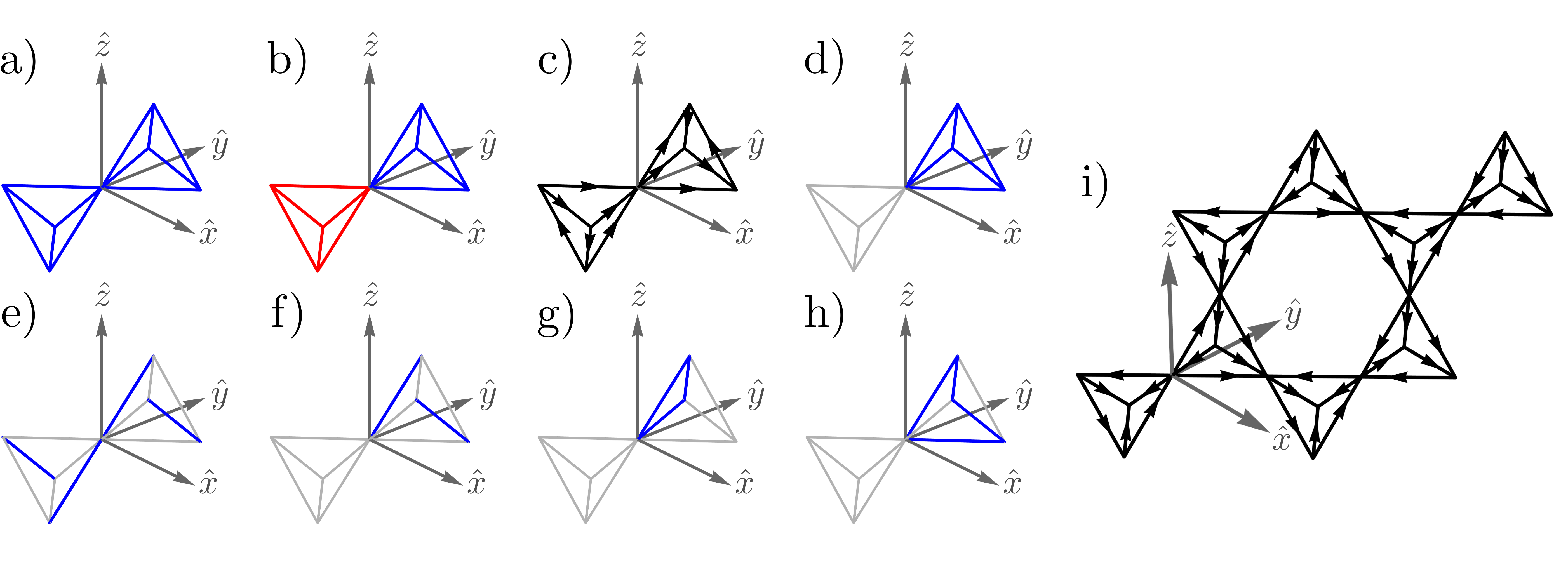}
\caption{Investigated hopping models on the pyrochlore lattice: The blue (red) bonds denote positive (negative) real-valued hoppings $ \chi$ ($- \chi$). Black arrows depict imaginary hoppings $i \chi$ ($-i \chi$) in (against) the direction of the arrow. Gray bonds carry zero hopping amplitudes. Within each ansatz, all finite hoppings have the same absolute value. a) The \textit{uniform} state. b) The \textit{staggered} state. c) The $(\pi/2,\pi/2,0)$ \textit{monopole} state. d) The localized \textit{tetramer} state breaking inversion symmetry. e) Two infinitely \textit{extended lines} breaking the $C_3$ rotation symmetry. f) Two localized \textit{dimers} breaking inversion and rotation symmetry. g) One localized \textit{trimer} breaking inversion and rotation symmetry. h) One localized \textit{fourfold loop} breaking inversion and rotation symmetry. i) The $(\pi/2,-\pi/2,0)$ \textit{monopole-antimonopole} state. \label{fig:HoppingAnsaetze}}
\end{figure*}

\begin{figure}[t]
\centering
\includegraphics[width=8.5cm]{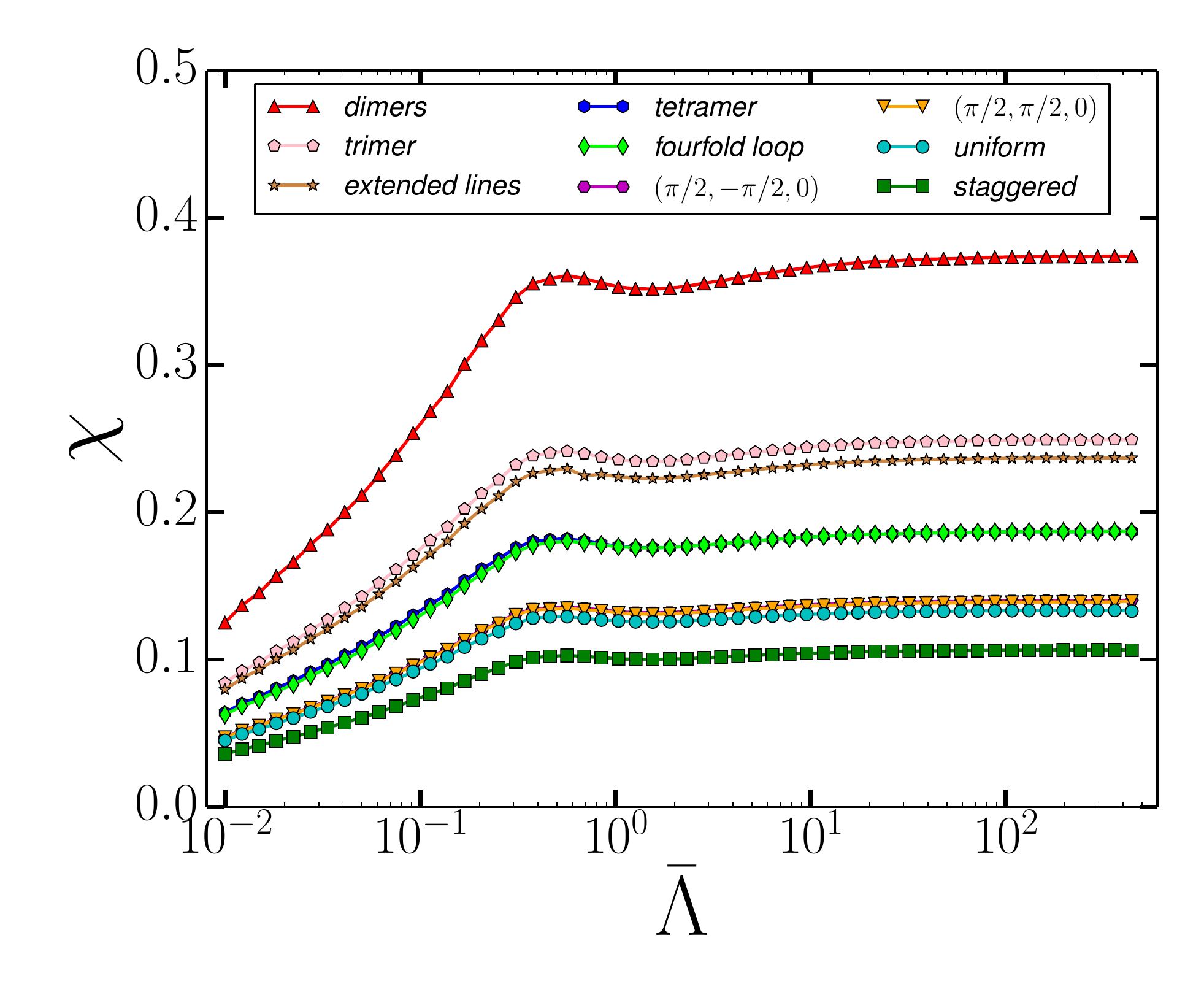}
\caption{Self-consistently determined nearest-neighbor hopping amplitudes as a function of the decoupling scale $\BL$ for the ans\"{a}tze given in Fig.~\ref{fig:HoppingAnsaetze}. While the amplitudes for symmetric quantum spin liquids are comparatively small, by far the largest amplitude results for the dimerized pattern in Fig.~\ref{fig:HoppingAnsaetze}(f). The decay of all amplitudes when $\BL\lesssim 0.4$ is caused by the finite pseudofermion lifetime due to the imaginary self energy from PFFRG. Note that the data points for the $(\pi/2,\pi/2,0)$ flux monopole state and $(\pi/2,-\pi/2,0)$ flux monopole-antimonopole state lie almost on top of each other, with slightly larger amplitudes for the $(\pi/2,-\pi/2,0)$ state. \label{fig:SCAmplitudes}}
\end{figure}

In this section, we present our results for the PFFRG-enhanced mean-field analysis for various ans\"{a}tze of the matrix $u_{ij}$. To simplify the analysis, we will restrict ourselves to finite nearest-neighbor hopping terms only, and thus all ans\"atze analyzed subsequently have a U(1) invariant gauge group (IGG)~\cite{Wen2002}. (Note that an ansatz with only hopping amplitudes is gauge equivalent to an ansatz with only real-valued pairing terms such that no bias is induced by concentrating on hopping ans\"atze.) We start by considering four simple or previously proposed candidate ans\"atze for symmetric and chiral quantum spin liquids which all satisfy the system's lattice symmetries projectively:
\begin{itemize}
    \item The \textit{uniform} ansatz with identical real-valued hoppings on all nearest-neighbor bonds, see Fig.~\ref{fig:HoppingAnsaetze}(a). Consequently, this state has a $0$-flux through the triangular faces of both up and down tetrahedra, and a $0$-flux through the elementary hexagon plaquettes of the pyrochlore lattice. This ansatz can be implemented within a four-site unit cell.
    \item The \textit{staggered} ansatz with real-valued and positive (negative) hoppings on all up (down) tetrahedra, see Fig.~\ref{fig:HoppingAnsaetze}(b). Such a pattern results in a $0$ ($\pi$)-flux through the triangular faces of up (down) tetrahedra, and a $\pi$-flux through the hexagons. This flux structure can be realized by a four-site unit cell.
    \item The $(\pi/2,\pi/2,0)$ flux \textit{monopole} state proposed in Ref.~\cite{Burnell2009} and depicted in Fig.~\ref{fig:HoppingAnsaetze}(c). This ansatz features a $\pi/2$-flux threading each of the triangular faces of both up and down tetrahedra, and a $0$-flux through the hexagons. Such a flux structure breaks inversion $(\mathcal{I})$ and time-reversal $(\mathcal{T})$ symmetries, but respects their product $\mathcal{IT}$, and thus describes a chiral spin liquid of the Kalmeyer-Laughlin type. It can be implemented within a four-site unit cell.
    \item The $(\pi/2,-\pi/2,0)$ flux \textit{monopole-antimonopole} state investigated in Ref.~\cite{Kim2008}. The ansatz features a $\pi/2$ $(-\pi/2)$-flux threading each of the triangular faces of up (down) tetrahedra, and a $0$-flux through the hexagons. This flux pattern involves a breaking of the screw $(\mathcal{S})$~\footnote{The screw symmetry here is a $\pi$-rotation about one of the three tetrahedral axis emanating out of the origin followed by a fractional translation by an amount which is equal to the length of the edge of the tetrahedra. It thus maps up and down tetrahedra into each other while preserving their orientation.} and time-reversal $(\mathcal{T})$ symmetries, while conserving their product $\mathcal{ST}$, and thus realizes a chiral spin liquid of a different symmetry class compared to the $(\pi/2,\pi/2,0)$ monopole flux state. This ansatz also involves a doubling of the unit cell along any two of the three Bravais lattice vectors, and thus requires a $16$-site unit cell, see Fig.~\ref{fig:HoppingAnsaetze}(i).
\end{itemize}

\begin{table*}
\setlength{\tabcolsep}{5pt}
\begin{tabular}[t]{lcllll}
Symmetry & $\eta$ & $(\Phi_{\bigtriangleup},\Phi_{\bigtriangledown},\Phi_{\hexagon})$ & Energy (VMC) &$\chi^{\BL\rightarrow 450 \, J}$ & $\chi^{\BL\rightarrow 0.01 \, J}$ \\ \hline
Fully symmetric & $+1$ & $(0,0,0)^{a}$~\cite{Kim2008,Burnell2009} & $-0.37502(6)$ &$0.1329$ &$0.0450$ \\ 
Fully symmetric & $+1$ & $(0,\pi,\pi)$ & $-0.37457(5)$ &$0.1063$&$0.0356$\\ 
Chiral $(\mathcal{IT})$ & $+1$ & $(\frac{\pi}{2},\frac{\pi}{2},0)^{b}$~\cite{Kim2008,Burnell2009} & $-0.457354(5)$&$0.1397$ &$0.0472$ \\
Chiral $(\mathcal{ST})$ & $-1$ & $(\frac{\pi}{2},-\frac{\pi}{2},0)^{c}$~\cite{Kim2008} & $-0.459402(6)$ &$0.1404$&$0.0475$\\ \hline \hline
\end{tabular}
\caption{The Gutzwiller projected ground state energies per site $E/J$ of the four U(1) ans\"atze obtained by variational Monte Carlo (VMC) calculations on a $6912$-site $(=4\times12\times12\times12)$ cluster which respects the full symmetry of the pyrochlore lattice. The ans\"atze with $\eta=1$ do not enlarge the four-site geometrical unit cell while the one with $\eta=-1$ involves a doubling of the unit cell along two tetrahedral axis, \textit{i.e.}, it has a $16$-site unit cell. The ans\"atze are completely characterized by specifying the gauge fluxes through three plaquettes, namely, the triangular faces of up tetrahedra $(\Phi_{\bigtriangleup})$, the triangular faces of down tetrahedra $(\Phi_{\bigtriangledown})$, and the hexagons $(\Phi_{\hexagon})$. The last two columns show the PFFRG-enhanced mean-field amplitudes at a large and a small decoupling scale $\BL\approx 450 \, J$ and $\BL\approx 0.01 \, J$.\\
$^{a}$ This Ansatz is labelled as $[0,0,0]$ in Ref.~\cite{Kim2008}, and referred to as the uniform state in Ref.~\cite{Burnell2009}.\\
$^{b}$ This Ansatz is referred to as the monopole flux state in Ref.~\cite{Burnell2009}. However, despite careful checking and benchmarking of our code, we find a different VMC energy [$E/J=-0.458525(4)$] of this state than given in Table I of Ref.~\cite{Burnell2009} [$E/J=-0.4473(9)$] on the $500$-site cluster employed therein. In Ref.~\cite{Kim2008}, this state is labelled as $\left [\frac{\pi}{2},\frac{\pi}{2},0 \right]$ and is referred to as the uniform flux state.\\
$^{c}$ In Ref.~\cite{Kim2008}, this Ansatz is referred to as the staggered flux state and labelled as 
$\left [\frac{\pi}{2},-\frac{\pi}{2},0 \right]$.\label{tab:VMC}}
\end{table*}

Restricting to these quantum spin liquid ans\"atze first, the results for the self-consistently obtained nearest-neighbor hopping amplitudes $\chi$ as a function of $\bar{\Lambda}$ are shown in Fig.~\ref{fig:SCAmplitudes}. At a pure mean-field level ($\bar{\Lambda}=\infty$), the two monopole states have the largest amplitudes, with a slight advantage for the $(\pi/2,-\pi/2,0)$ state, in confirmity with the Rokhsar rules~\cite{Rokhsar-1990,Kim2008}. Interestingly, this preference for the monopole states remains qualitatively unchanged as $\bar{\Lambda}$ is lowered (\textit{i.e.}, as more effects of renormalization are taken into account), a fact which is also corroborated by Gutzwiller projection of the corresponding mean-field states, cf. Appendix~\ref{app:VMC}, whereby the two monopole type states give the lowest variational energies, with a slightly lower energy for the $(\pi/2,-\pi/2,0)$ monopole-antimonopole state of Ref.~\cite{Kim2008}, in agreement with the findings from PFFRG~\footnote{In Tab.~I of Ref.~\cite{Kim2008}, it was found that the energy of the $(\pi/2,-\pi/2,0)$ state is higher compared to the energy of the $(\pi/2,\pi/2,0)$ state. However, we find this to be a finite-size effect as their calculations were done on a $4\times4^{3}=256$-site cluster. Our calculations on a $4\times12^{3}=6912$ site cluster show that, in fact, the energy of the $(\pi/2,-\pi/2,0)$ state is lower.}, see Table~\ref{tab:VMC}. The overall size of amplitudes, however, is renormalized to smaller values below $\bar{\Lambda}\approx0.4$ which is generally expected when incorporating fluctuations beyond mean field. On a technical level, this is a consequence of the imaginary on-site self-energy $\gamma^\Lambda(\omega)$ in PFFRG which acts as a finite lifetime for the pseudofermions. Surprisingly, the hierarchy of the amplitudes and their ratios remain largely unchanged upon varying $\bar{\Lambda}$. 
In total, these results indicate that on the level of spin liquid states, previous results finding a preference for the chiral flux states is confirmed, with our study indicating a slight preference for the monopole-antimonopole state (in contrast to previous studies), and this property seems relatively robust with respect to renormalization effects.

Next, we extend our analysis by including various ans\"atze which explicitly break the system's lattice symmetries, see Figs.~\ref{fig:HoppingAnsaetze}(d)-(h). These states restrict the pseudofermions to be localized on \textit{dimers}, \textit{trimers}, \textit{tetramers}, \textit{fourfold loops}, and extended \textit{1D lines}. The ans\"atze can be characterized by the lattice symmetries they break, particularly inversion $i$ which transforms $\vec{r}\rightarrow-\vec{r}$ (where the origin coincides with a pyrochlore lattice point) and $C_3$ rotation which performs a $2\pi/3$ rotation around an axis connecting the midpoints of two adjacent tetrahedra (this axis passes through the pyrochlore lattice point common to both tetrahedra). The pattern in Fig.~\ref{fig:HoppingAnsaetze}(d) [Fig.~\ref{fig:HoppingAnsaetze}(e)] then breaks only inversion $i$ [only $C_3$ rotation] while the ans\"atze in Fig.~\ref{fig:HoppingAnsaetze}(f)-(h) break both $i$ and $C_3$. Note that the dimer and tetramer states have been recently discussed in Refs.~\cite{Neupert2021,hagymasi21} and both works find a tendency for their realization in the ground state of the pyrochlore Heisenberg antiferromagnet. Our results in Fig.~\ref{fig:SCAmplitudes} likewise show that the dimer state yields by far the largest amplitudes throughout the entire range of $\bar{\Lambda}$ and also outperforms the spin liquid ans\"atze. One can interpret this as an indication for dimerization, however, we explicitly stress that at this stage of the analysis one needs to be careful with this conclusion. Particularly, our approach, while unbiased when comparing amplitudes of different quantum spin liquids among each other may be biased with respect to finding explicit symmetry breaking. Our large intra-dimer hopping amplitudes $\chi_\text{dimer}$ certainly imply that dimer formation leads to a significant reduction of energy {\it on the dimer bonds}. However, all other `non-dimer' bonds may be energetically unfavorable, which is not captured in our approach such that dimerization may appear more favorable than it actually is (as mentioned before, the total energy would serve as an ultimate diagnostic measure). Please confer Sec.~\ref{sec:discussion} for a more detailed discussion.
A definite conclusion on possible dimerization cannot be drawn solely based on these results. We therefore, consult a different approach in the next section.

\subsection{Investigation of symmetry breaking patterns}\label{sec:Results_symm_breaking}
\begin{figure}[t]
\centering
\includegraphics[width=8.5cm]{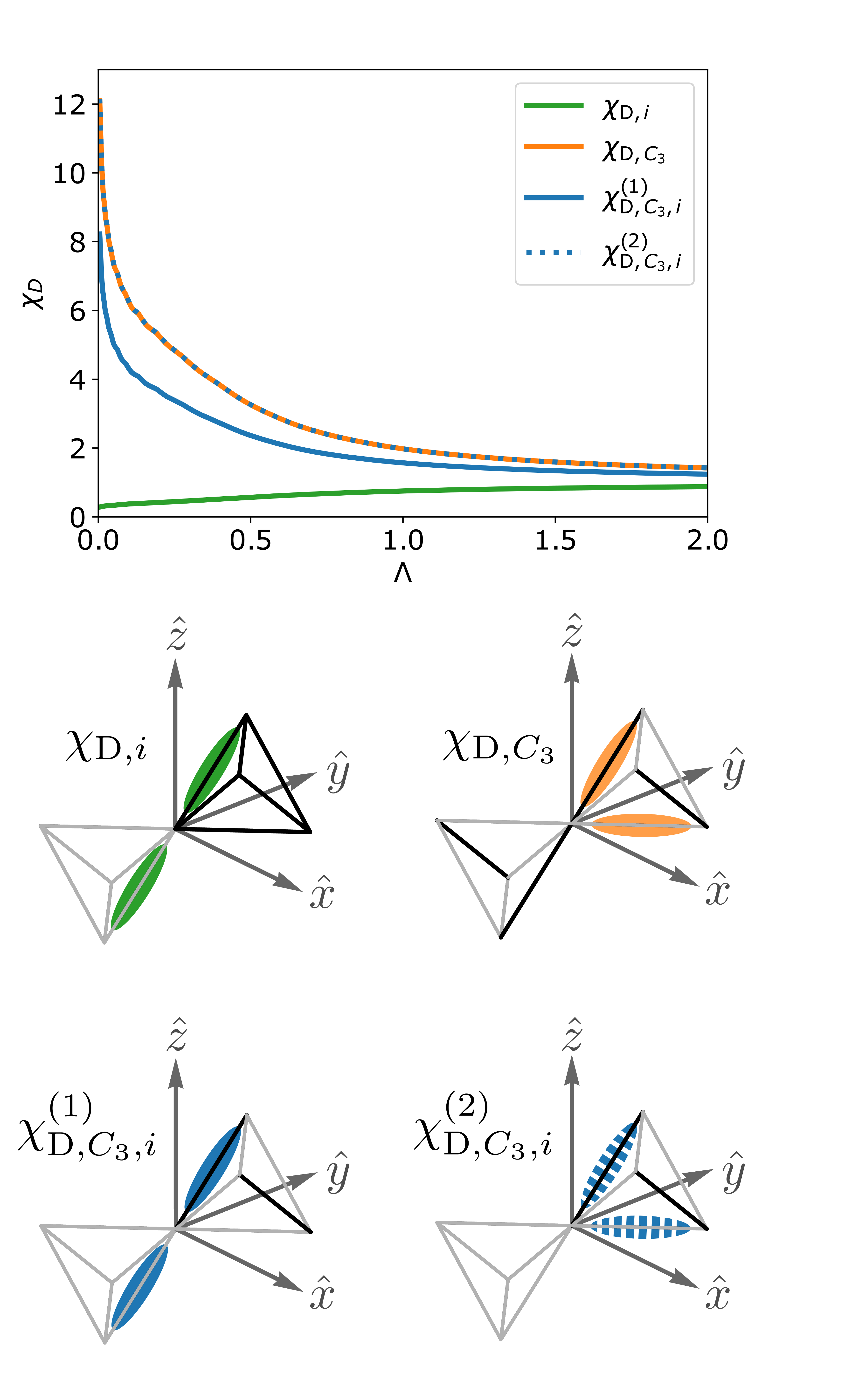}
\caption{Renormalization group flows of response functions $\chi_D$ [see Eq.~\eqref{eq:dimer_response}] for the three symmetry breaking perturbations illustrated in Fig.~\ref{fig:HoppingAnsaetze}(d), (e), and (f). In the bottom part of the figure, the colored bonds illustrate which weakened (gray) and strengthened (black) bonds are compared to each other (the colors in these illustrations match the ones of the curves). Note that for the dimer pattern in Fig.~\ref{fig:HoppingAnsaetze}(f) two different response functions $\chi_{\text{D},C_3,i}^{(1)}$ and $\chi_{\text{D},C_3,i}^{(2)}$ can be defined, depending on the precise choice of weakened and strengthened bonds that are compared (bottom part of the figure).\label{fig:dimerCorrelations}}
\end{figure}

We continue probing the Hamiltonian in Eq.~\eqref{eq:Hamiltonian} with respect to various symmetry breaking perturbations using the approach explained in Sec.~\ref{sec:symm_break}. Particularly, we impose the patterns of Figs.~\ref{fig:HoppingAnsaetze}(d), (e), and (f) by strengthening the blue bonds and weakening the other ones. This probes the system with respect to inversion symmetry breaking only [see Fig.~\ref{fig:HoppingAnsaetze}(d)], $C_3$ rotation symmmetry breaking only [see nematic pattern in Fig.~\ref{fig:HoppingAnsaetze}(e)] and a combination of both [see dimer pattern in Fig.~\ref{fig:HoppingAnsaetze}(f)].

Note that for the patterns in Figs.~\ref{fig:HoppingAnsaetze}(d) and (e) all weak bonds are symmetry equivalent (\textit{i.e.}, they can all be mapped onto each other by applying symmetry transformations of the remaining {\it unbroken} symmetries) and the same is true for the strong bonds. Hence, there is a unique way of defining the corresponding response functions $\chi_{\text{D},i}$ and $\chi_{\text{D},C_3}$ [see Eq.~\eqref{eq:dimer_response}] by comparing the spin correlations on weakened and strengthened bonds. On the other hand, for the dimer pattern in Fig.~\ref{fig:HoppingAnsaetze}(f), the weakened bonds cannot all be mapped onto each other due to the reduced number of point symmetries. Therefore, there are two distinct possibilities to compare the spin correlations on weakened and strengthened bonds, the inter-tetrahedron response $\chi_{\text{D},C_3,i}^{(1)}$ and the intra-tetrahedron response $\chi_{\text{D},C_3,i}^{(2)}$, which are defined as illustrated in Fig.~\ref{fig:dimerCorrelations}.

A first important observation is that the response function $\chi_{\text{D},i}$ which probes pure inversion symmetry breaking decreases as $\Lambda$ flows towards zero. This clearly indicates that the system rejects this pattern, which is in agreement with the relatively small hopping amplitudes found in our PFFRG-enhanced parton approach. In stark contrast, but again in agreement with our previous analysis, the nematic and dimer response functions $\chi_{\text{D},C_3}$, $\chi_{\text{D},C_3,i}^{(1)}$ and $\chi_{\text{D},C_3,i}^{(2)}$ exhibit a pronounced increase at small $\Lambda$, where the initial perturbation gets amplified by roughly one order of magnitude. As expected, $\chi_{\text{D},C_3,i}^{(1)}\neq\chi_{\text{D},C_3,i}^{(2)}$ since there is no symmetry relation connecting them. We find that the intra-tetrahedron response $\chi_{\text{D},C_3,i}^{(2)}$ is larger, indicating the system's strong propensity for symmetry breaking already within one tetrahedron. Interestingly, $\chi_{\text{D},C_3,i}^{(2)}$ and $\chi_{\text{D},C_3}$ are numerically indistinguishable such that our results are compatible with both patterns. Based on this analysis we conclude that the system shows tendencies for the realization of either a dimer or a nematic ground state.

\section{Discussion}\label{sec:discussion}
Several comments on the applied methods and our results are in order. As explained before, the PFFRG-enhanced mean-field approach may have a bias towards detecting symmetry breaking dimerization patterns. Despite this possible bias, we believe that the large dimer amplitude in Fig.~\ref{fig:SCAmplitudes} still points towards a valence bond solid formation. Our investigation of symmetry breaking perturbations in Sec.~\ref{sec:Results_symm_breaking} substantiates this conclusion. Another indication comes from calculating and comparing dimer amplitudes for other well-studied frustrated models such as Heisenberg models on the square lattice (with first and second-neighbor couplings) and on the kagome lattice. For these two systems we observe similar trends of overpredicting dimer states. However, dimerization tendencies are found to be significantly stronger on the pyrochlore lattice than for the square and kagome lattice systems. Particularly, taking the amplitude $\chi_\text{uniform}$ of a uniform hopping model (equal real-valued hopping amplitudes on all nearest-neighbor bonds) as a reference, we find that the ratio $r=\chi^{\bar{\Lambda}\rightarrow 0}_\text{dimer}/\chi^{\bar{\Lambda}\rightarrow 0}_\text{uniform}$ is largest for the pyrochlore lattice: $r_\text{pyrochlore}\sim 2.8$, $r_\text{square}\sim 2.4$, $r_\text{kagome}\sim 2.2$ (where for the square lattice a ratio between second and first neighbor antiferromagnetic interactions of $\frac{J_2}{J_1}=\frac{1}{2}$ is assumed and for the kagome lattice only nearest-neighbor couplings are considered). 

Before we move on to our conclusions, it is worth discussing the PFFRG-enhanced parton approach from a more general perspective. Instead of interpreting it as an improvement of a bare mean-field theory, one can take the opposite viewpoint and consider it as a simplification of a rather sophisticated `itinerant spinon FRG'. In this latter scheme the spinon degrees of freedom would be treated more explicitly by allowing the fermions to hop already on the level of the FRG (which implies making a gauge fixing). This, however, makes the numerical efforts vastly more complicated, since one can no longer benefit from the simplifying locality of free fermionic propagators (one possibility to proceed is to sacrifice frequency resolution for including itinerant spinons~\cite{roscher19}). Starting from an `itinerant spinon FRG' our PFFRG-enhanced parton approach would be obtained by restricting spinon hopping only to the fermionic particle-hole ladder channel while neglecting it in all other channels (diagrammatic contributions without any spinon hopping processes would still be treated in all FRG channels). This is because singling out one decoupling channel is equivalent to a mean-field treatment in this channel, therefore restoring our PFFRG-enhanced parton approach. With these properties, our technique is well adapted to describe non-magnetic ground states of quantum spin models, including quantum spin liquids, but at the same time avoids the difficulties associated with the explicit description of itinerant fractional excitations. In the present implementation, however, the ground state is identified as the state with the largest mean-field amplitudes $u_{ij}$. This criterion is borrowed from a bare mean-field approach [see Eq.~(\ref{eq:EMF})] and could be improved in more advanced schemes where the minimization of the free energy is taken as a diagnostic tool. We defer this to future work as it requires more in-depth method development.

\section{Conclusion}\label{sec:conclusion}

In this work, we have investigated the ground state properties of the nearest-neighbor $S=1/2$ pyrochlore Heisenberg antiferromagnet using principally the PFFRG method and variants thereof. We have first introduced and discussed in detail the PFFRG-enhanced parton mean-field technique on which parts of our analysis are based. This method uses a parton ansatz for the system's spinon degrees of freedom and determines the parameters of the corresponding bilinear parton Hamiltonian (spinon hopping and pairing amplitudes) self-consistently. Compared to a standard parton mean-field treatment, our approach makes explicit use of renormalized vertices from PFFRG and, hence, includes important effects of quantum fluctuations not contained at the bare mean-field level. As an improvement of an earlier implementation in Ref.~\cite{hering19}, we have further resolved a technical difficulty associated with an overcounting of fermionic diagrams.

The PFFRG-enhanced parton mean-field technique is primarily designed to probe the spin liquid nature of frustrated quantum spin systems. Applying it in this context first, we have found that, among the previously proposed spin liquid ans\"atze, the monopole-antimonopole state of Ref.~\cite{Kim2008} is the preferred one, with the monopole flux state of Ref.~\cite{Burnell2009} being a close competitor, and finally the symmetric spin liquids are found to be noncompetitive. Interestingly, apart from an overall reduction of spinon amplitudes, fluctuations beyond mean-field turn out to have a rather small effect. The hierarchy of spin liquids has been validated by our large-scale variational Monte Carlo calculations which find the Gutzwiller projected wave function energies of the two chiral spin liquids as the lowest, with the monopole-antimonopole state having a slightly lower energy $[E/J=-0.459402(6)]$ compared to the monopole state $[E/J=-0.457354(5)]$, while the two symmetric spin liquids are found to have comparatively higher energies, $E/J=-0.37502(6)$ and $E/J=-0.37457(5)$ for the $(0,0,0)$ and $(0,\pi,\pi)$ flux states, respectively. It is also worth mentioning that these energies are considerably larger than the ground state energies found in Refs.~\cite{Neupert2021,hagymasi21} and references therein, providing an independent argument against a spin liquid ground state.

In the next step, we have applied our PFFRG-enhanced parton mean-field theory to ans\"atze which explicitly break the system's lattice symmetries mimicking valence bond solid formations. The spinon amplitudes for dimer patters are found to be dominant and even outperform the ones of our spin liquid ans\"atze. While we have argued that due to a possible methodological bias this result should not be overinterpreted as an indication for dimerization, a comparison to square and kagome models still suggests that dimer tendencies are particularly strong on the pyrochlore lattice. This observation is underpinned more rigorously within a direct investigation of symmetry breaking patterns where PFFRG has been applied to model Hamiltonians with slightly weakened or strengthened bonds according to the dimerization pattern to be probed. With this technique we have identified a clear tendency for either $C_3$ rotation symmetry breaking or a combination of both $C_3$ and inversion symmetry breaking. On the other hand, inversion symmetry breaking alone is clearly not supported. In total, all these results indicate that the ground state of the nearest-neighbor antiferromagnetic Heisenberg model on the pyrochlore lattice is either given by a dimer or a lattice nematic state.

We have exclusively concentrated on a single model without varying any coupling parameters. However, exploring how the ground state of this model is embedded in a wider phase diagram remains an interesting problem. For example, by interpolating between the Heisenberg and the Ising model the system will undergo a phase transition to a U(1) spin liquid at an unknown Ising interaction strength. Studying the fate of the putative symmetry broken dimer/nematic state upon adding longer-range Heisenberg couplings constitutes another possible future research direction. Since dimer states show the largest energy reduction on bonds occupied by a dimer, longer-range couplings on bonds without dimers may be energetically unfavorable in a valence bond solid. The associated destabilization of symmetry breaking states may induce quantum spin liquid behavior and possibly realize the monopole-type states which we found to be the preferred quantum spin liquid ans\"atze in the nearest-neighbor model.

From a methodological perspective, we have demonstrated the applicability of the PFFRG-enhanced parton mean-field theory to complex quantum spin models and showed its capability to smoothly interpolate between a bare mean-field scheme and a fully one-loop renormalized approach. The surprisingly small effect of the renormalized vertices which only amounts to an overall reduction of mean-field amplitudes raises questions about whether the system has an intrinsic mean-field character or whether the current level of renormalization is insufficient to have a more significant impact on the spinon amplitudes. Since our PFFRG-enhanced parton mean-field approach is formulated in a very general way and can be based on arbitrary types of renormalized vertex functions (as long as a diagrammatic overcounting is prevented), plenty of possibilities for improvements are opened up. Particularly interesting would be the use of vertex functions from the recently developed multiloop schemes~\cite{thoenniss20,Kiese2020_2} which could yield further insight into the ground state properties of frustrated quantum spin systems.

\section{Acknowledgments}
We thank Chunxiao Liu, Leon Balents, and Federico Becca for illuminating discussions on fermionic spin liquid ans\"atze.
Y.\,I. acknowledges financial support by Science and Engineering Research Board (SERB), Department of Science and Technology (DST), India through the Startup Research Grant No.~SRG/2019/000056, MATRICS Grant No.~MTR/2019/001042, and the Indo-French Centre for the Promotion of Advanced Research (CEFIPRA) Project No. 64T3-1. This research was supported in part by the  National Science Foundation under Grant No.~NSF~PHY-1748958, the Abdus Salam International Centre for Theoretical Physics (ICTP) through the Simons Associateship scheme funded by the Simons Foundation, IIT Madras through the Institute of Eminence (IoE) program for establishing the QuCenDiEM group (Project No. SB20210813PHMHRD002720) and FORG group (Project No. SB20210822PHMHRD008268), the International Centre for Theoretical Sciences (ICTS), Bengaluru, India during a visit for participating in the program “Novel phases of quantum matter” (Code: ICTS/topmatter2019/12). Y.\,I. acknowledges the use of the computing resources at HPCE, IIT Madras. F.\,F. acknowledges support from the Alexander von Humboldt Foundation through a postdoctoral Humboldt fellowship. 
V.\,N. would like to thank the HPC service of ZEDAT, Freie Universität Berlin, for computing time. J.R. acknowledges financial support by the German Research Foundation within the CRC 183 (project A04). M.H. thanks the Department of Physics, Freie Universität Berlin, for computing time at the tron cluster and, in particular, J\"org Behrmann for outstanding IT support.

\appendix

\section{Numerical solution of the Bethe-Salpeter equation}\label{app:BSE}

In this appendix, we explain how we solve the Bethe-Salpeter equation [Eq.~\eqref{eq:BetheSalpeter} from the main text]. In real space, the convolution integral over internal momenta turns into a direct product 
\begin{equation}
    \tilde{\Gamma}^{\Lambda}_{\vk-\vq} - \int\limits_{\mathrm{BZ}} \frac{\dd\vec{p}}{V_{\mathrm{BZ}}} \, \tilde{\Gamma}^{\Lambda}_{\vk-\vec{p}}\Gamma^{\Lambda}_{\vec{p}-\vq} \rightarrow  \tilde{\Gamma}^{\Lambda}_{ij} -  \tilde{\Gamma}^{\Lambda}_{ij}  \Gamma^{\Lambda}_{ij} \, .
\end{equation}
This decouples the different real-space components of $\Gamma^{\Lambda}$ yielding one Fredholm integral equation of the second kind in Matsubara space for each component. Numerically, we deal with discretized Matsubara frequencies $\omega_{\kappa}$ and the required vertex functions depend on a single frequency argument. We abbreviate $\Gamma^{\Lambda}(\omega_{\kappa},-\omega_{\kappa},0)\rightarrow\Gamma^{\Lambda}(\omega_{\kappa})$ from now on and rewrite the integral equations as
\begin{equation}\label{eq:DiscreteBetheSalpeter}
    \Gamma^{\Lambda}_{ij}(\omega_{\kappa})=\tilde{\Gamma}^{\Lambda}_{ij}(\omega_{\kappa})+ \sum\limits_{\kappa'}W^{\Lambda}_{ij}(\omega_{\kappa},\omega_{\kappa'})\tilde{\Gamma}^{\Lambda}_{ij}(\omega_{\kappa'}),
\end{equation}
where $W^{\Lambda}_{ij}(\omega_{\kappa},\omega_{\kappa'})$ are weights to approximate the continuous integral in Eq.~\eqref{eq:BetheSalpeter}. We assume a locally parabolic behavior of the integrand which, for equidistant frequencies, would be identical to using Simpson's numerical integration rule~\cite{abramowitz+stegun}. In the discretized Matsubara space, Eq.~\eqref{eq:DiscreteBetheSalpeter} is a matrix equation and can be solved for the desired two-particle irreducible vertex
\begin{equation}\label{eq:BSEsolution}
    \tilde{\Gamma}^{\Lambda}_{ij}(\omega_{\kappa})=\sum\limits_{\kappa'}\left[\mathds{1}+W^{\Lambda}_{ij} \right]^{-1} (\omega_{\kappa}, \omega_{\kappa'})\Gamma^{\Lambda}_{ij}(\omega_{\kappa'}).
\end{equation}
The inverse of $\mathds{1}+W^{\Lambda}_{ij}$ can be computed efficiently from a lower-upper decomposition. The resulting vertex function $\tilde{\Gamma}^{\Lambda}_{ij}(\omega_{\kappa})$ is then plugged into the self-consistent Fock equation [Eq.~\eqref{eq:FRGDecoupling} from the main text].

\section{Computation of self-consistent mean-field amplitudes}\label{app:SCSol}

Here we discuss how we numerically solve the self-consistency equation of our PFFRG-enhanced parton mean-field approach for a given ansatz $u_{ij}$. The left-hand side (LHS) and the right-hand side (RHS) of Eq.~\eqref{eq:FRGDecoupling} have identical structures in sublattice and momentum space if $u_{ij}$ is inserted from a PSG classification. This is also true for the local and quasi one-dimensional hopping patterns presented in Figs.~\ref{fig:HoppingAnsaetze}(d)-(h). Therefore, it is sufficient to evaluate a specific component of $u^{\BL}_{\vk}$ in momentum and sublattice space which is proportional to a single free mean-field parameter $\xi$. By singling out one amplitude, the integrand on the RHS of Eq.~\eqref{eq:FRGDecoupling} should only contain the Fourier transform of the according vertex function due to the real-space structure of contributing Feynman diagrams [see Fig.~\ref{fig:BSE}(c)], \textit{i.e.}, the vertex function $\tilde{\Gamma}^{\bar{\Lambda}}_{\vk-\vec{q}}$ only contains the Fourier transform of the nth-neighbor vertex function if an nth-neighbor hopping or pairing term is considered.

For an efficient momentum integration, the Brillouin zone is chosen to be cuboidal and we compute the RHS integrand on an equidistant and symmetric mesh containing $(n_k)^3$ points in momentum space. In Matsubara space, the integrand is computed for $n_{\omega}$ symmetric points between $-\Lambda_{\text{max}}$ and $\Lambda_{\text{max}}$ where we call the largest (infrared) cutoff scale used within the PFFRG framework $\Lambda_{\text{max}}$. The frequency points include the discrete mesh points $\omega_{l}$ from PFFRG, which are more densely distributed in the ultraviolet $\omega\rightarrow 0$ limit, if $|\omega_{l}-\omega_{l+1}|<2\Lambda_{\text{max}}/n_{\omega}$ and for equidistant frequencies otherwise. We also insert $\omega=0$ as well as $k_{\mu}=0$ with $\mu \in \{x,\,y,\,z \}$ to our grids such that we have an odd number of points on all integration axes. On these frequency and momentum grids, the inverse of the matrix $i\omega+i\gamma^{\bar{\Lambda}}(\omega) - u^{\bar{\Lambda}}_{\vec{q}} $ is evaluated with the builtin matrix inversion of \textit{numpy}. The integration is then carried out via Simpson's integration in \textit{scipy}. 

Typically, we compute the RHS at two different points in parameter space $\xi=\xi_{-}$ and $\xi=\xi_{+}=\xi_{-}+\Delta_{\xi}>\xi_{-}$ yielding the two values $f(\xi_{-})$ and $f(\xi_{+})$. We use an iterative scheme in which we start with $\xi_{-}=0$ and $\xi_{+}=\Delta_{\xi}$ as well as some fixed $n_{\omega}$ and $n_k$. If the conditions
\begin{equation}\label{eq:SolutionCondition}
    f(\xi_{-})>\xi_{-} \quad \text{and} \quad f(\xi_{+})<\xi_{+} \, 
\end{equation} 
are fulfilled, we find a solution of the self-consistency equation at 
\begin{equation}
    \xi_{\text{sol}}=\frac{f(\xi_{-})\Delta_{\xi}-\xi_{-}\left (f(\xi_{+})-f(\xi_{-})\right)}{\Delta_{\xi}-\left(f(\xi_{+})-f(\xi_{-})\right)} \, ,
\end{equation}
where the two lines connecting either $\xi_{-}$ and $\xi_{+}$ or $f(\xi_{-})$ and $f(\xi_{+})$, respectively, intersect. If the conditions in Eq.~\eqref{eq:SolutionCondition} are not met, we repeat the previous procedure after increasing $\xi_{\pm}$ by $\Delta_{\xi}$~\footnote{At least if we consider a single amplitude, the RHS of the self-consistency equation always fulfills $f(-\xi)=-f(\xi)$. In this case, it is sufficient to only consider positive amplitudes without loss of generality.}. 

In all simulations, we keep the number of frequencies $n_{\omega}$ sufficiently large and, in particular, larger than the number of frequencies at which the vertex functions are computed within PFFRG. If we find a solution it will thus still depend on the discretization parameters $\Delta_{\xi}$ and $n_k$. In order to eliminate these dependencies, we repeat the above procedure for a given $n_k$ after sending $\Delta_{\xi}\rightarrow\Delta_{\xi}'=\Delta_{\xi}/2$. The solution is considered to be converged if
\begin{equation}\label{eq:ConvergenceCondition}
 |\xi_{\text{sol}}(\Delta_{\xi}',n_k')-\xi_{\text{sol}}(\Delta_{\xi},n_k)|<\delta \xi \, ,
\end{equation}
where $\delta \xi$ is absolute error that we allow for. Once convergence is reached in $\Delta_{\xi}$, we repeat this procedure after increasing $n_k\rightarrow n_k'= n_k+2$ until Eq.~\eqref{eq:ConvergenceCondition} is also fulfilled for $n_k$. For all results presented in this work, we set $\delta \xi=10^{-4}$.

\section{Variational Monte Carlo}\label{app:VMC}

As discussed in Sec.~\ref{sec:PFFRG}, the parton representation of spin operators introduced by Eq.~\eqref{eq:Arbikosov} is accompanied by an artificial enlargement of the Hilbert space, with the inclusion of unphysical $S=0$ fermionic states with doubly-occupied and/or empty sites. As a consequence, mean-field wave functions defined within the pseudofermionic framework do not represent well-defined quantum states for the original spin Hamiltonian. A way to overcome this drawback and define appropriate variational states for the spin problem is constraining the pseudofermionic wave functions to the spin Hilbert space. This can be achieved by the application of the Gutzwiller projector, 
\begin{equation}
\mathcal{P}_{\text{G}}=\prod_i (f^\dagger_{i\uparrow}f_{i\uparrow}-f^\dagger_{i\downarrow}f_{i\downarrow})^2,
\end{equation}
to the mean-field state $|\Phi_{\text{MF}}\rangle$. The resulting wave function, 
\begin{equation}\label{eq:variationalwf}
|\Psi_{\text{var}}\rangle=\mathcal{P}_{\text{G}}|\Phi_{\text{MF}}\rangle,
\end{equation}
can be employed as a faithful variational state for the Heisenberg Hamiltonian [Eq.~\eqref{eq:Hamiltonian}]. The Gutzwiller projection can be easily implemented within a Monte Carlo scheme where the sampling of the mean-field state is limited to the configurations with one fermion per site. In Tab.~\ref{tab:VMC}, we report the variational energies obtained by Gutzwiller-projecting the mean-field spin-liquid ans\"atze defined in Sec.~\ref{sec:Results_partons}.
\\

\bibliography{refs1}

\end{document}